\def\year{2015}
\documentclass[letterpaper]{article}
\usepackage{aaai}
\usepackage{amsmath}
\usepackage{times}
\usepackage{booktabs}
\usepackage{helvet}
\usepackage{graphicx}
\usepackage{courier}
\DeclareMathOperator*{\argmax}{arg\,max}

\newtheorem{definition}{Definition}

\newtheorem{example}{Example}

\newtheorem{principle}{{Principle}}

\usepackage{hyperref}
\newcommand{\tabincell}[2]{\begin{tabular}{@{}#1@{}}#2\end{tabular}}
\newcommand{\footremember}[2]{%
	\footnote{#2} 
	\newcounter{#1}
	\setcounter{#1}{\value{footnote}}%
}
\newcommand{\footrecall}[1]{%
\footnotemark[\value{#1}]%
}
\newcommand{\nop}[1]{}

\begin{document}
\title{Long Concept Query on Conceptual Taxonomies}
\author{Yi Zhang\footremember{yi} \\, Yanghua Xiao\footremember{yh} \\, Seung-won Hwang\footremember{swh} \\, Wei Wang \footremember{ww} \\ \\
\footrecall{yi} \footrecall{yh} Fudan University, Shanghai, China \\
\footrecall{swh} Yonsei Uninversity, Seoul, Korea \\
\footrecall{ww} University of California, Los Angeles, CA, USA\\
\footrecall{yi}\footrecall{yh} \{z\_yi11, shawyh\}@fudan.edu.cn, \footrecall{swh}seungwonh@yonsei.ac.kr, \footrecall{ww}weiwang@cs.ucla.edu\\
}

\maketitle
\begin{abstract}
This paper studies the problem of finding typical entities when the concept is given as a query. For a short concept such as \emph{university}, this is a well-studied problem of retrieving \emph{knowledge base} such as Microsoft's Probase and Google's isA database pre-materializing entities found for the concept in
Hearst patterns of the web corpus. However, we find most real-life queries are long concept queries (LCQs), such as \emph{top American private university}, which cannot and should not be pre-materialized.
Our goal is an online construction of entity retrieval for LCQs. We argue a naive baseline of rewriting LCQs into an intersection of an expanded set of composing short concepts leads to highly precise results with extremely low recall. Instead, we propose to augment the concept list, by identifying related concepts of the query concept. However, as such increase of recall often invites false positives and decreases precision in return, we propose the following two techniques:
First, we identify concepts with different relatedness to generate linear orderings and pairwise ordering constraints. Second, we rank entities trying to avoid conflicts with these constraints, to prune out lowly ranked one (likely false positives).
With these novel techniques, our approach significantly outperforms state-of-the-arts.     

\end{abstract}


\section{Introduction}
We study the problem of returning entities for the name of a concept as a query. For example, a user may query {\it top American private university} to obtain the most typical examples such as \{\textsl{Harvard, Stanford, Yale}\} as results. 

\begin{figure}
\centering
\includegraphics[scale=0.45]{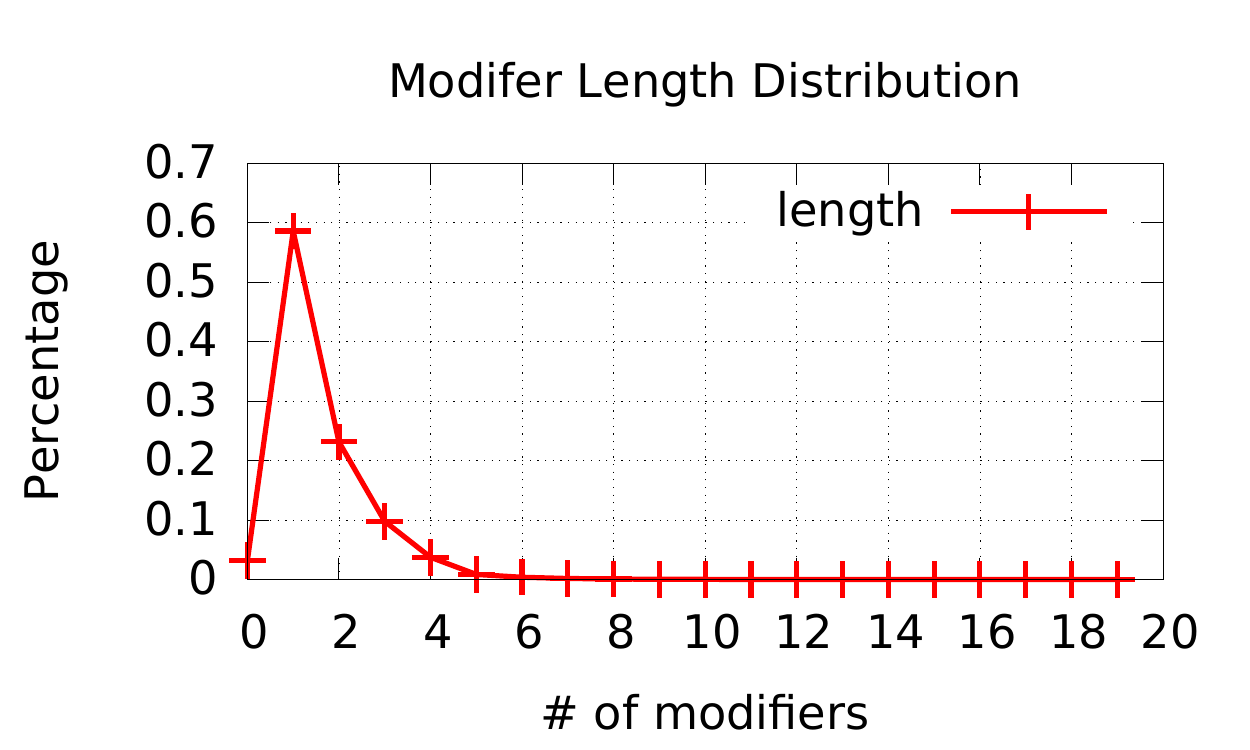}
\vspace{-3mm}
\caption{\small{Modifier length distribution in Probase}}
\vspace{-6mm}
\label{fig:dis}
\end{figure}

One related effort is retrieving \emph{knowledge base} materializing isA relationships between entity-concept pairs, such as Microsoft's Probase and Google's isA database. 
These knowledge bases are constructed by extracting Hearst patterns from web corpus, such as `American universities such as Harvard and Yale'.
However, Figure~\ref{fig:dis} shows the distribution of
{\it \# of modifiers} used in the concepts of Probase,
indicating that more than 60\% of concepts have zero or one modifier.
In other words, this resource cannot answer long concept query (LCQ)
such as {\it top American private university}, which is dominant in search engine queries.
A statistical study reveals that LCQ accounts for a majority of searches on search engine: based on a one week worth of search log (from 07/25/2012 to 07/31/2012) of BING~\cite{wang2014head}, about 90\% of distinct queries have one or more modifiers which differs significantly from Figure~\ref{fig:dis}.
Our goal is to bridge such a gap, by an online materialization for LCQs, using the materialized concepts in Probase.~\footnote{Note we do not prematerialize all LCQs, as materialization of
concepts of length $n$ generates $2^{n-1}$ combinations.}

Another related effort is \emph{set expansion}.
In this problem, a small set of seed entities are given as a query, such as \{\textsl{Harvard, Yale}\}, to obtain more instances such as \emph{Stanford}.
As we empirically show later, our solution can solve this task, by identifying related concepts to the query and return the typical examples, but we cannot use existing set expansion solutions for our problem, as they require example entities which directly reveal the query intent as input.
Existing work on set expansion has many applications-- a) knowledge base expansion to mine more instances and b) automatic list completion in spreadsheet.
Our work, by automatically finding possible sets of seed entities, can apply to the above and more,
enabling also the completion of the concept/list with few or even no examples.

Toward this goal, we first describe a high-precision baseline and discuss the challenges in improving recall while maintaining precision.

\subsubsection*{High-precision Baseline}
A naive baseline is decomposing the given long concept $c_q$, e.g., \emph{top American private university}, into a set of composing shorter concepts $\mathbf{C}_q$, e.g., \{\emph{top university, American university, private university}\}.
We can intersect the instances of $c \in \mathbf{C}_q$. However, as knowledge bases are extracted from text patterns demonstrating a few examples (and not aiming to enumerate all instances),
the recall of instances in each concept is low and this problem exacerbates as more concepts are intersected.
For example, the concept {\it low-fertility country} in Probase contains only 4 entities:\{\textsl{Japan, Germany, Spain, Canada}\}. However, none of them is among the top 20 lowest fertility countries in 2014. Many typical instances in this concept such as \textsl{Singapore} and \textsl{South Korea} are missing due to the rare mentions by Hearst patterns.
 
We thus propose our approach to solve this problem and the main idea is to first, increase the recall by introducing new concepts to $\mathbf{C}_q$. For example, adding \emph{Ivy League}, which is nearly equivalent to the query concept \textsl{top American private university}, to $\mathbf{C}_q$ would significantly increase the recall.
However, some of such addition of a concept may introduce false positive instances, for which we devise ranking for pruning.

\subsubsection*{Recall}
We have two basic ideas to improve the recall, corresponding to two situations for the instance intersection of shorter concepts: $\cap_{c\in \mathbf{C}_q}e(c)$ is empty or not, where $e(c)$ is the instances of $c$ found from the knowledge base. When it is not empty, which implies that we have found a certain number of seed instances. These seeds are strong signals for us to find more related concepts and in turn to find more result entities. However, in most cases, $\cap_{c\in \mathbf{C}_q}e(c)$ is empty due to the sparsity of current knowledge bases. In these cases, we find more sets of seed instances by relaxing the constraint on the seed instances. Specifically, instead of using $\cup_c\in {\mathbf{C}_q}$ as seeds, we use $\cap_{c\in \mathbf{C}'_q}e(c)$ such that $\mathbf{C}'_q\subset \mathbf{C}_q$, i.e., the {\it instance intersection of subsets of shorter concepts}. Since we intersect less concepts, it is quite possible to find a $\mathbf{C}'_q$ whose intersection is not empty. Given non-empty intersections, we use them as seed   
and expand the concept in the same way as the previous case.

\nop{
We introduce concepts based on different kinds of examples to generate two kinds of ordering constraints of entities respectively to improve the recall of LCQs. As for the examples of LCQ, we can find the concepts nearly equivalent to $c_q$ which means related to all concepts in $\mathbf{C}_q$ that can generate a new permutation of entities. This permutation is a global ordering constraints which may help us optimize the entity ranking. Another kind of examples is the example from a subset of $\mathbf{C}_q$ which can generate concepts related to some aspects of $c_q$. Based on the importance of the examples, we can get that some of the concepts are more likely to be equivalent to $c_q$, thus the entities of these concepts are more likely to be the answers than the entities of other concepts. We consider these entity list pairs as pairwise ordering constraints which may help us optimize the entity ranking locally.   
}
\subsubsection*{Precision}
After we get all candidate entities, we still need to rank them together. In the concept expansion step, we have derived a linear ordering and pairwise ordering constraints.  We need to aggregate all these information including an entity ordering derived without introducing any expanded concept to generate a better global ranking with fewer conflicts. Thus, we propose a new aggregation method based on a learning-to-rank framework to solve this problem, combined with both global and pairwise constraints.

\smallskip
To conclude, our problem is that given a long query concept $c_q$, return its top-$k$ entities which are the $k$ most typical instances of $c_q$. Our framework is as follows.
First, generate an entity ranking list for the query concept without introducing some other concepts by an iterative method. It provides us a baseline ranking for entities bias to the precision which is an important global ordering constraint. 
Second, we propose to find possible seed instances for the LCQ, and propose an instance-based concept expansion to improve the recall. Here, we expand concepts by two probabilistic models, Noisy-Or model and Naive-Bayes model.  
Third, we generate two kinds of constraints based on the baseline ranking and the concepts expanded. A new entity ranking aggregation method is proposed to minimize the conflicts of the constraints. The new ranking can maintain a high precision while improving the recall.

\section{Related Work}

Retrieving typical instances of the given concept was studied, often in the context
 of building a knowledge base. Microsoft's Probase and Google's isA database mine textual patterns, such as Hearst patterns, to collect isA relations of concepts.
  However, long concepts are sparsely observed in the text and thus cannot be obtained in this manner.
  Our problem distinguishes itself for being \emph{online process} of mining instances for \emph{LCQ}.
We consider baselines of leveraging existing knowledge bases materializing short concepts:
First, we can decompose a LCQ into composing short concepts and intersect the instances, which leads to highly precise but very sparse results.
Second, the above set can be expanded by \emph{set expansion} work, which we describe below.

Set expansion considers that a small set of seed entities are given as a query, such as \{\textsl{Harvard, Yale}\}, to obtain more instances.
The approach is to take the collection (seed entities) as input to find related entities to populate the input set.

Google Sets is a product implementation used to populate a spreadsheet after users provide some instances as examples. Inspired by Google Sets, many research works followed~\cite{bayesian,seisa,seal2,more,pantel2009web}, to measure the membership strength of an item to the hidden concepts exemplified by the query entities.

Related problems include harvesting lists on the Web, and retrieving the list that completes the seed instances~\cite{seisa,wangchi}. Compared with these works, our task is more challenging relying solely on entities and concepts in the knowledge base without rich exclusive lists from the Web. Most of the entity lists of the concept in the knowledge base are incomplete and inaccurate which makes the inference much more difficult.

However, this line of approaches may lead to ambiguity when given a long concept query. For example, when we get seed entities \{\textsl{Yale, Harvard, Stanford}\} as input from the decomposition of the concept \textsl{American private university}, we lose the information that the results we need should not only belong to American universities but also private universities. Set expansion approaches tend to mix entities belonging to different concepts during expansion which will lead to a lot of false-positives.

Our proposed solution can complement both knowledge base construction and set expansion.
For knowledge base, we can use our solution to mine instances for the missing concepts or grow
the instance set of the concept that is too sparse.
For set expansion, our approach allows to consider instance lists, by aggregating them with the awareness of the
semantic relationships of concepts.

Another line of related work is to aggregate different rankings. It is also named as Kemeny rank aggregation~\cite{k1,k2,k3,k4}, consensus ranking~\cite{consensus}, median ranking~\cite{medians},and preference aggregation~\cite{preference}. These ranking aggregation methods have been of particular interest in the information retrieval and database communities. All these works are defined on a set of permutations. However, in our problem, we not only need to aggregate linear orderings (permutations) but also some pairwise ordering constraints which are not permutations.



\section{Baseline Ranking}
In this section, we present our baseline ranking model.
This model underlies that our further development in the next two sections. 


The baseline ranking is based on the following mutually recursive principles:
\begin{principle}
An entity $e$ is unlikely to be the answer, if it is related to none of the concepts in $\mathbf{C}_q$.
\end{principle}

\begin{principle}
A concept $c$ in $\mathbf{C}_q$ is not informative, if it contains none of the entities which are likely to be the answer.
\end{principle}

The rationality of the principles are as follows. Due to the data sparsity of the knowledge base, an entity is not very likely to be related to every concept in $\mathbf{C}_q$ even it is one of the answers. We assume that only if the entity is not related to any of the concept in $\mathbf{C}_q$, then the entity is not the answer. Hence, based on our principles, an answer entity does not have to be related to every concept to get a high probability.  Similarly, the concept is {\it informative} (promising to be the concept whose entities are the answers) if it contains at least one entity which is very likely to be the answer. In general, the more important entities the concept contains, the more informative the concept is. 

Clearly, the Noisy-Or model can reflect the above principles and rationalities. Here, we use $\sigma(e)$ to represent the probability of the entity $e$ to be the answer, and $\sigma(c)$ to represent the informativeness of the concept $c$. We have:
\begin{equation}
\small
\sigma(e)=1-\prod_{c\in c(e), c\in \mathbf{C}_q}{(1-\sigma(c))}
\label{eq:sig1}
\end{equation}  

\begin{equation}
\small
\sigma(c)=1-\prod_{e\in e(c)}{(1-\sigma(e))}
\label{eq:sig2}
\end{equation}
where $c(e)$ is the concepts of $e$ found from the knowledge base.

In Eq.~\ref{eq:sig1} and Eq.~\ref{eq:sig2}, $1-\sigma(c)$ and $1-\sigma(e)$ are usually quite small and multiplying many of them may lead to underflow. To facilitate computation, we use logarithm to transfer the equations into their equivalent logarithmic forms:
\begin{equation}
\small
log(1-\sigma(e))=\sum_{c\in c(e), c\in \mathbf{C}_q}{log(1-\sigma(c))}
\end{equation}

\begin{equation}
\small
log(1-\sigma(c))=\sum_{e\in e(c)}{log(1-\sigma(e))}
\end{equation}

As in Authority-hub analysis~\cite{ah} and PageRank~\cite{pagerank}, we adopt an iterative method to compute $\sigma(c)$ and $\sigma(e)$. We choose the initial state in which all concepts $c\in \mathbf{C}_q$ have a uniform probability to be an informative concept. In each iteration, our algorithm first uses $\sigma(c)$ to compute $\sigma(e)$, and then recomputes $\sigma(c)$ based on the $\sigma(e)$. The iteration stops when it reaches a stable state where for each $e\in E$, $\sigma(e)$ changes a little after an iteration~\cite{truth}.  \nop{\bf .. guarantee that the algorithm will teminate at a steady state.... give referece...}


\section{Concept Expansion}

In this section, we introduce how we improve the recall by finding more related concepts. 
We first show how we expand concept by instances when $E(\mathbf{C}_q)=\cap_{c\in \mathbf{C}_q}e(c)\neq \emptyset$.
For the cases where $E(\mathbf{C}_q)=\emptyset$, we find a subset $\mathbf{C}'_q$ of $c_q$ so that its instance intersection is not empty.

\subsection{Instance-based Expansion}

We first present our principle for instance-based concept expansion in Principle 3 and illustrate it in Example~\ref{exp:1}. 
There are two conditions in the principle. Next, we elaborate each one. 

In the first condition, we claim that a strong signal of semantic equivalence or relatedness between concept pairs is their overlap of respective instances. Thus, if a concept's entities contain most of the entities in $\cap_{c\in \mathbf{C}_q} e(c)$, it is very likely that the concept is a closely related concept of $c_q$. As shown in Example~\ref{exp:1}, the concept \textsl{Ivy League} contains almost all instances in  $\cap_{i=1}^{3}e(c_i)$.

\nop{
In this subsection, we assume that the overlap of all the shorter concepts of the query exists which is the simplest situation. Formally $\cap_{c\in \mathbf{C}_q}e(c)\neq \emptyset$, which means example instances of $c_q$ can be found, we can expand some nearly equivalent concepts of $c_q$ based on these examples. 
}

\nop{Here, we make an assumption that $\cap_{c \in \mathbf{C}_q}{e(c)}$ is not empty. For those $c_q$ that $\cap_{c \in \mathbf{C}_q}{e(c)}$ is empty, we generate pairwise ordering constraints which are described in the next subsection.}

\begin{principle}
A concept $c$ is semantically related to $c_q$, if (1) most instances in $ \cap_{c\in \mathbf{C}_q} e(c)$ are related to $c$ and (2) $c$ has as few as possible instances $e\notin \cup_{c\in \mathbf{C}_q} e(c)$
\end{principle}

\begin{example}
Suppose $c_q$: {\it top American private university}, then $\mathbf{C}_q$=\{\textsl{$c_1$: top university, $c_2$: American university, $c_3$: private university}\}. In Probase, we have $\cap_{i=1}^{3} e(c_i)$ =\{\textsl{harvard, princeton, yale, pennsylvania state university, cornell}\}. Clearly, \textsl{`Ivy League'} is a closely related concept of $c_q$. In Probase, it contains most of the entities in $\cap_{i=1}^{3}e(c_i)$ and few entities not in  $\cup_{i=1}^{3}e(c_i)$. Hence,  we can use instances in \textsl{Ivy League} such as \{\textsl{dartmouth, columbia}\} to improve the recall.  
\label{exp:1}
\end{example}

By the second condition, we hope to exclude an over-general concept.
Theoretically, if a concept contains many entities not related to any shorter concept of $c_q$, it is not a good concept to expand. In our running example, another concept \textsl{famous university} contains all entities in $\cap_{i=1}^{3}e(c_i)$. However, it also contains entities such as \textsl{loughborough}, which is not a member of any of shorter concepts. Thus, \textsl{famous university} should not be considered.

Formally, we define a relevance score $rel(c,\mathbf{C}_q)$ to rank the concepts expanded which reveals that how likely $c$ is to be a semantically related concept of $c_q$. Clearly, the Naive-Bayes model can reflect our principle. The model is as follows:
\begin{equation}
\small
\begin{aligned}
rel(c, \mathbf{C}_q)=\frac{P(c|E(\mathbf{C}_q))}{g(c)} \propto \frac{P(E(\mathbf{C}_q)|c)P(c)}{g(c)}\\
=\frac{P(c)\prod_{e \in E(\mathbf{C}_q)}P(e|c)}{g(c)}
\end{aligned}
\end{equation} In the nominator, $\prod_{e \in E(\mathbf{C}_q)}P(e|c)$ implies that only when most instances in $E(\mathbf{C}_q)$
are related to $c$, $c$ will get a high ranking score.

Generally, there are relatively few concepts related to all of the entities in the intersection due to the sparsity in the current knowledge base. Therefore, appropriate smoothing is necessary to avoid zero probabilities. To do this, we can assume that with probability $(1-\gamma)$, the concept should be chosen by its prior typicality.
\begin{equation}
\small
=\frac{P(c)\prod_{e\in E(\mathbf{C}_q)}{(\gamma P(e|c)+(1-\gamma)P(e))}}{g(c)}
\label{eq:2}
\end{equation}

Another option to attack the sparsity problem is relaxing the first condition in the Principle 3 by a Noisy-Or model:
\begin{equation}
\small
rel(c, \mathbf{C}_q)=\frac{P(c|E(\mathbf{C}_q))}{g(c)}
 =\frac{1-(1-\lambda)\prod_{e\in E(\mathbf{C}_q)}{(1-P(c|e))}}{g(c)}
\label{eq:1}
\end{equation}
where $\lambda$ is the leak probability that the concept $c$ is a related concept of $c_q$ for no good reasons. The nominator is the probability that $c$ is a concept of $E(c_q)$ and the denominator is a function $g(c)$ to penalize the extent containing noisy entities. By Noisy-Or model, any concept that is related to at least one entity in $E(c_q)$ will be captured.

In the above equations, we still need to estimate $P(c|e)$ (the probability that $c$ is a concept of $e$) and $P(e|c)$ (the probability that $e$ is an instance of $c$) as well as the prior probability of $e$ and $c$. We use Probase data for the estimation as follows:
\begin{equation}
\small
P(c|e)=\frac{n(c,e)}{n(e)}, P(e|c)=\frac{n(c,e)}{n(c)}, 
\end{equation}
where $n(c,e)$ is the number of co-occurrence between $c$ and $e$, and $n(e)$ is the number of the occurrence of $e$ in Probase. We set $P(e)\propto n(e)$ and $P(c)\propto n(c)$.

There are many functions that can be used to define $g(c)$ varying from different data quality of the knowledge base, and we use the following one:
\begin{equation}
\small
g(c)=\frac{\delta+\sum_{e\in e(c), e\notin E_{\cup}(\mathbf{C}_q)}{(n(e,c)+1)}}{\sum_{e\in e(c)}(n(e,c)+1)}
\label{eq:g}
\end{equation}
where $E_{\cup}(\mathbf{C}_q)=\cup_{c'\in \mathbf{C}_q} e(c')$ and $\delta$ is small positive value less than 1. In the denominator, we count all entities that do no appear in instances of any shorter concept of $c_q$.

\paragraph*{Entity Ordering}
After the instance-based expansion, we derive the relevance score $rel(c, \mathbf{C}_q)$.
Given these scores, we rank all entries as follows:
\begin{equation}
\small
rel(e, c_q)=\sum_{c}{P(e|c)rel(c,\mathbf{C}_q)}
\label{eq:rank}
\end{equation} The rationality of the ordering is obvious. The more relevant concepts contain an entity, the more likely the entity is a good answer. Next, we highlight two facts about the ordering.
First, {\it $rel(e, c_q)$ generates a linear ordering on all observed entities in the intersection as well as expanded entities}.
Although theoretically we cannot exclude the possibilities of ties, it is a rare case practically. Even tie happens, we can assign a random ordering to ensure the result ordering is linear.
Second, {\it the ordering is well defined on all concepts, including those in $\mathbf{C}_q$}. In general, for a $c\in \mathbf{C}_q$, Eq~\ref{eq:g} has
a smallest value with the nominator as $\delta$ because $c$ contains no entity outside of  $E_{\cup}(\mathbf{C}_q)$.
Hence, we assign almost no penalty to  $c\in \mathbf{C}_q$ and consequently these concepts have a high
relevance score. This is intuitively correct.

\nop{
Then it is easy to generate a ranking of entities based on their relatedness to the equivalent concepts which is a global ordering constraint for the results. The computation is as follows:
\begin{equation}
rel(e, c_q)=\sum_{c}{P(e|c)rel(c,\mathbf{C}_q)}
\label{eq:rank}
\end{equation}
}

\subsection{Seed Instances Generation}
For the cases where $E(\mathbf{C}_q)=\emptyset$, our basic idea is to find a subset $\mathbf{C}'_q$ of $\mathbf{C}_q$ such that $E(\mathbf{C}'_q)$ is not empty.
For each $E(\mathbf{C}'_q)\neq \emptyset$, we use the instance-based expansion to find new entities.
Continue Example~\ref{exp:1}, all shorter concept subsets $\{c_1\}$, $\{c_2\}$, $\{c_3\}$, $\{c_1, c_2\}$, $\{c_1, c_3\}$, and $\{c_2, c_3\}$ are candidate $\mathbf{C}'_q$. Since we intersect less concepts, we have more chance to find a non-empty $E(\mathbf{C}'_q)$. 
 
More formally, given a long concept $c_q$, let $n$ be the number its modifiers, which means that $|\mathbf{C}_q|=n$. In our implementation, we exhaustively enumerate all $\mathbf{C}'_q$ with capacity less than $n$. 
For most LCQs, we have $2^n-2$ choices of $\mathbf{C}'_q$. In general, $n$ is less than 10, hence the enumeration cost is acceptable. 

\paragraph*{Pairwise Ordering Constraint}
Note that the seed instance generation inherently implies a pairwise ordering constraint.
Clearly, the more shorter concepts join the intersection,
the more likely their concepts and entities expanded are related to the query concept.
 If $e_1$ has a higher order than $e_2$, we denote it as 
$e_1\succ e_2$. We formalize the constraint in Def~\ref{def:po}.
We illustrate the rationality of the constraint in Example~\ref{exa:cons}.
These constraints actually represent the partial order between entities. 
\begin{definition}[Pairwise Ordering Constraint]
For any $e_i\in E(\mathbf{C}^{(1)}_q)$ and $e_j\in E(\mathbf{C}^{(2)}_q)$ such that $\mathbf{C}^{(1)}_q\subset \mathbf{C}_q$ and $\mathbf{C}^{(2)}_q\subset \mathbf{C}_q$, if $|\mathbf{C}^{(1)}_q|> |\mathbf{C}^{(2)}_q|$, we have $e_i\succ e_j$.
\label{def:po}
\end{definition}

\nop{
Here $E'(\mathbf{C}^{(1)}_q)$ and $E'(\mathbf{C}^{(2)}_q)$ refer to the expanded entity set of $\mathbf{C}^{(1)}_q$ and $\mathbf{C}^{(2)}_q$ respectively.}

\begin{example}
Continue the running example, \textsl{`great state school'} contains entities \{\textsl{virginia, unc, michigan, berkley}\} in Probase which overlaps only with $c_1$ and $c_2$. Instead \textsl{`Ivy League'} overlaps with $c_1$, $c_2$ and $c_3$. Hence, it is reasonable to believe that instances of  \textsl{`Ivy League'} are more likely to be the correct answers than that in \textsl{`great state school'}. In other words, we have the following constraint $\{harvard, princeton,...\succ berkley, virginia,...\}$  
\label{exa:cons}
\end{example}

\section{Ranking Aggregation}

In this section,  we present our solution to aggregate all ranking of entities under the constraints. 
We propose a probabilistic ranking optimization method based on Bradley-Terry model. This model quantifies the probability of generating a ranking permutation, and our goal is to identify the most likely permutation under the given constraints.

More specifically, we need to generate a better linear ordering by aggregating the baseline ranking ($R_b$) and ranking of entities after concept expansion ($R_c$) and the pairwise ordering constraints ($R_p$).
We first derive our problem model by Bradley-Terry Model.

\subsection{Bradley-Terry Model}

 Given a pair of individuals $i$ and $j$, it estimates the probability that the pairwise comparison $i>j$ turns out true, as
\begin{equation}
\small
P(i>j)=\frac{\pi_i}{\pi_i+\pi_j}
\end{equation}
where $\pi_i$ and $\pi_j$ is a positive real-value score assigned to individual $i$ and $j$.
Our goal is to find an ideal scoring $S$ that maximizes the probability of a permutation.
The probability can be computed from both a global and a partial ordering constraint.
Given $S$, $s_x$ and $s_y$  represent the ranking score of $x$ and $y$, and  $\pi_x=e^{s_x}$.
Then, the probability that $x\succ y$ is:
\begin{equation}
\small
P(x\succ y)=\frac{e^{s_x}}{e^{s_x}+e^{s_y}}
\end{equation}

We now discuss how to formulate such probability for 
\begin{itemize}
\small
\item Total ordering $Z=\{Z_i\}_{i=1}^{n-1}$, where $Z_i=z_i\succ z_{i+1}, z_{i+2} ..., z_n$ as follows:
\begin{equation}
\small
P(Z_i)=P(z_i\succ \{z_j|j>i\})=\frac{e^{s_{z_i}}}{e^{s_{z_i}}+\sum_{j>i}{e^{s_{z_j}}}}
\end{equation}
Then we denote $L(Z)$ as the probability to generalize the total ordering:
\begin{equation}
\small
L(Z)=log(\prod_{i=1}^{n-1}P(Z_i))=\sum_{i=1}^{n-1}log(P(Z_i))
\end{equation}
\item Pairwise ordering between $X=\{{x_i}\}$ and $Y=\{{y_j}\}$, i.e.,  $X\succ Y$ as follows:
\begin{equation}
\small
P(X\succ Y)=\frac{\sum_{x_i\in X}{e^{s_{x_i}}}}{\sum_{x_i\in X}{e^{s_{x_i}}}+\sum_{y_j\in Y}{e^{s_{y_j}}}}
\end{equation}
Then the log likelihood of all partial order constraints in $R_p$, i.e., $L(R_p)$ can be computed as follows:
\begin{equation}
\small
L(R_p)=\sum_{(X_i,Y_i)\in R_p}log(P(X_i\succ Y_i))
\end{equation}
\end{itemize}

Thus, our goal is finding a scoring function $S$ to maximize the probability obtained from total orderings (from baseline ranking $R_b$ and entity ranking after concept expansion $R_c$) as well as the partial orders (from seed instance generation $R_p$). Hence, our ranking aggregation problem is:
\begin{equation}
\small
\argmax_S{(1-\alpha -\beta )L(R_b)+\alpha L(R_c)+ \beta L(R_p)}
\end{equation} where $\alpha, \beta$ are parameters used to tune the relative importance of each part. There are some strategies to tune the parameters, but they are out of the range of this paper. In our implementation, we just assign the same weight to all the parameters. 

\subsection{Model Computation}
Since $S$ (the scores of all entities) is a set of parameters which we need to learn, we use stochastic gradient descent~\cite{sgd} to compute the model.
The gradient of every part of our optimization function are as follows:
\begin{equation}
\small
\frac{\partial L(R_b)}{\partial s_i}=1-\frac{e^{s_i}}{\sum_{j=i}^{n}e^{s_j}}, \frac{\partial L(R_c)}{\partial s_i}=1-\frac{e^{s_i}}{\sum_{j=i}^{n}e^{s_j}}
\end{equation}
\begin{equation}
\small
\frac{\partial L(R_p)}{\partial s_i}=\sum_{(X_j,Y_j)\in R_p, X_j\ni i}(\frac{e^{s_i}\sum_{k\in Y_j}e^{s_k}}{\sum_{k\in X_j}e^{s_k}(\sum_{k\in X_j}e^{s_k}+\sum_{k\in Y_je^{s_k}})})
\end{equation}


\section{Evaluation}
In this section, we evaluate the precision and recall of our approaches with the comparison with the baselines and some of the state-of-the-art approaches.

\begin{figure}
\centering
\includegraphics[scale=0.325]{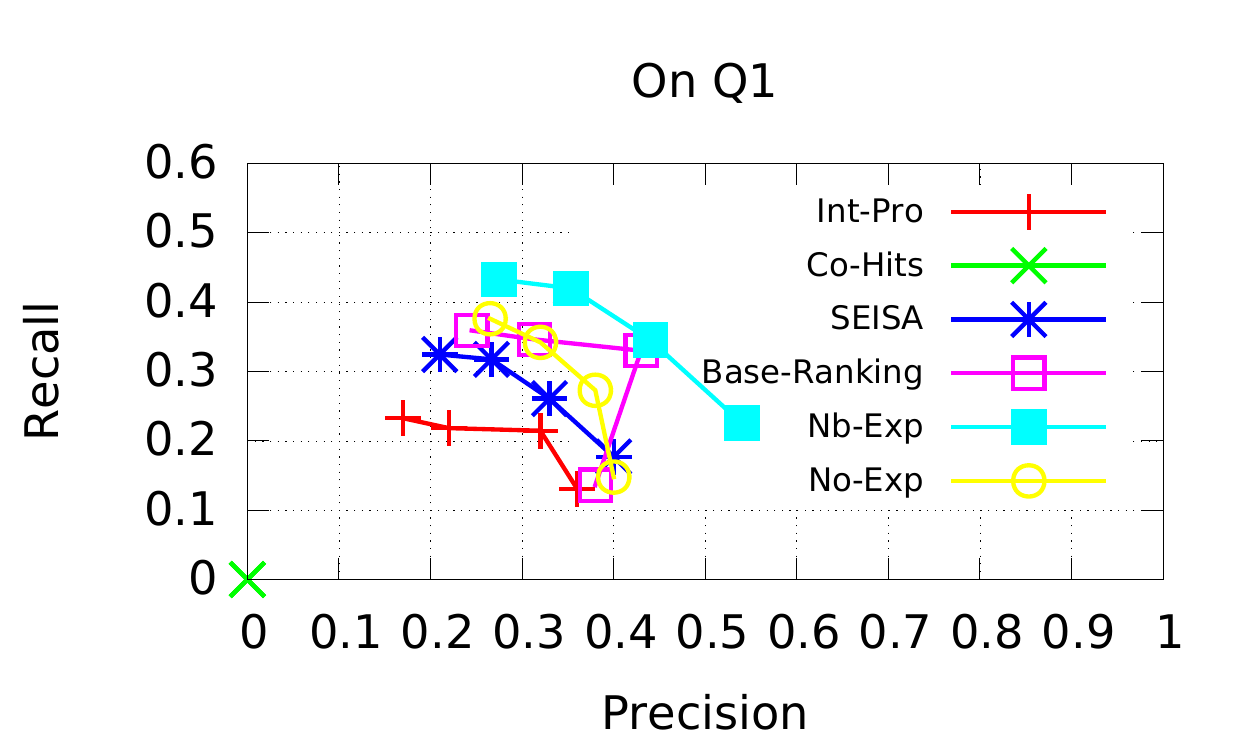}
\includegraphics[scale=0.325]{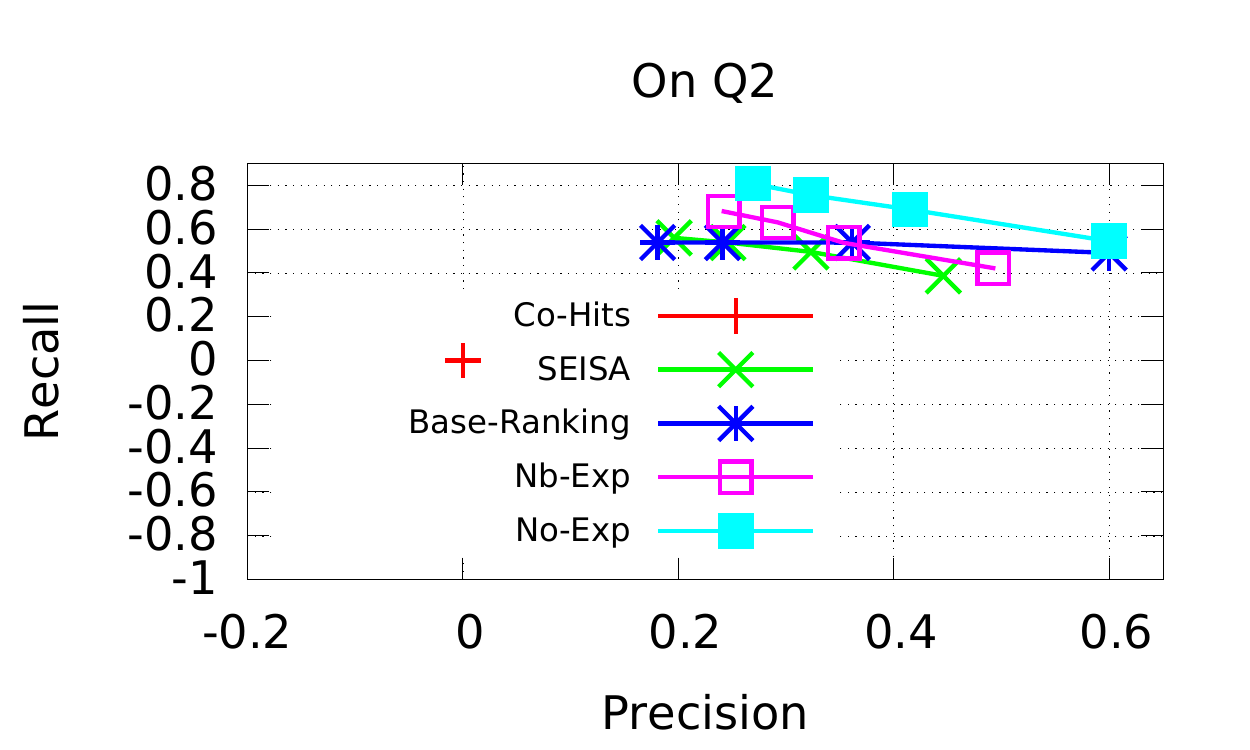}
\vspace{-7mm}
\caption{\small{Precision and Recall curve $@k$}}
\vspace{-5mm}
\label{fig:wp}
\end{figure}

\subsection{Experimental Setup}

We use Probase, which has a high coverage over entities and concepts, as our knowledge repository from which we retrieve answers for long concept query. To construct the long query set, we first find all concepts consisting of two words whose head occurs in more than four different concepts in Probase. Thus, for each head, we get many different modifiers. We randomly choose part of modifiers (from two to four) and combine them with the head to construct a long concept. We manually check each randomly generated long concept. If it is a meaningful concept, we accept it otherwise reject it.

\subsubsection{Dataset}
We generate two long query sets, a real set and a synthesized set respectively to evaluate our approach. 
\begin{itemize}

\item \textbf{Real data set $Q_1$.}~
\textbf{$Q_1$} consists of 15 long concepts which we can find ground truth answer from Wikipedia directly. If the answer entity list exists in Wikipedia, we directly use it as the ground truth answer. Otherwise, we first find the entity lists of all shorter concepts in $\mathbf{C_q}$ and use their intersections as the ground truth answer. If any shorter concept has no entity list in Wikipedia, we just skip the long concept. 
By comparing the results generated by our solution with the ground truth answer, we can directly test the effectiveness of our solution. 

\item \textbf{Synthetic data set $Q_2$.}~
Due to the data sparsity, entities in Probase may not overlap with some of the entities in Wikipedia and vice versa. This leads to a low recall and precision of our query answering. Hence, we further construct a synthesized query data set $Q_2$.
This query set consists of 20 long concepts. For these long concepts or their shorter concepts, we can not find the corresponding entity lists from Wikipedia. Yet the entity intersection of shorter concept is not empty in Probase. Then, we randomly eliminate half of the entities from the intersection and test whether our solution can find these deliberately removed entities.
\end{itemize}

\begin{figure}
\centering
\includegraphics[scale=0.325]{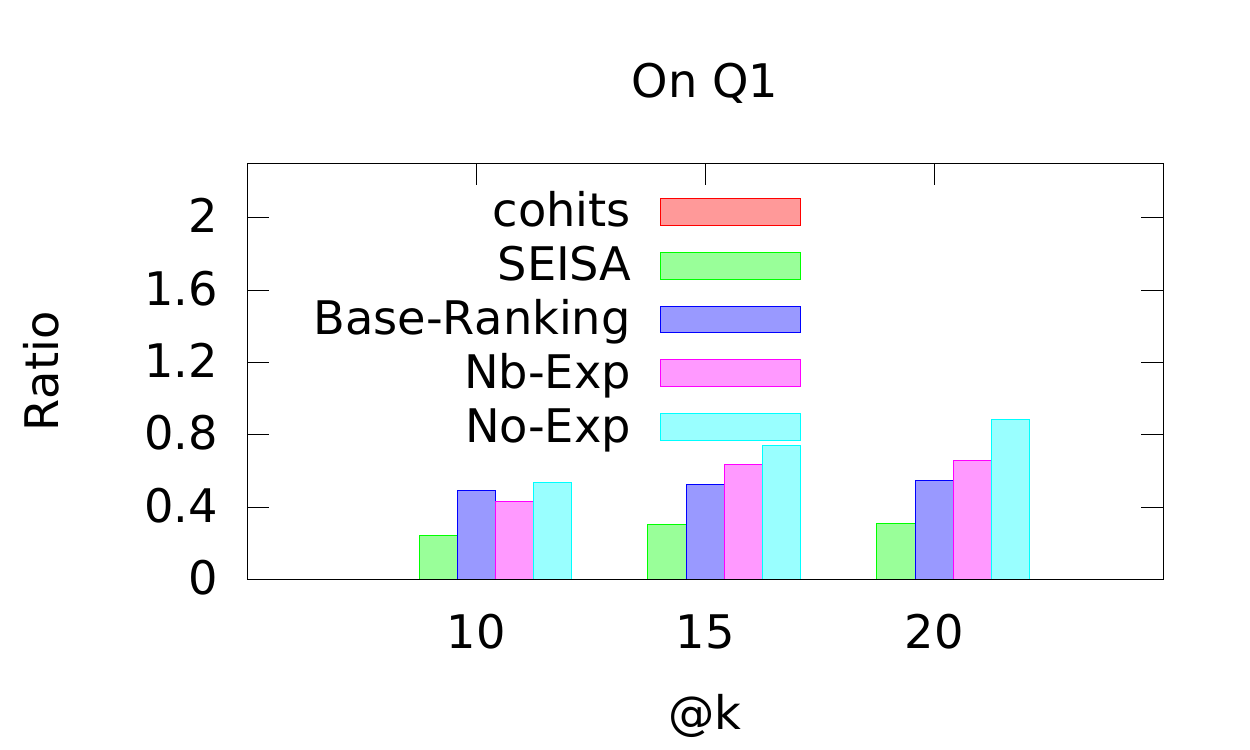}
\includegraphics[scale=0.325]{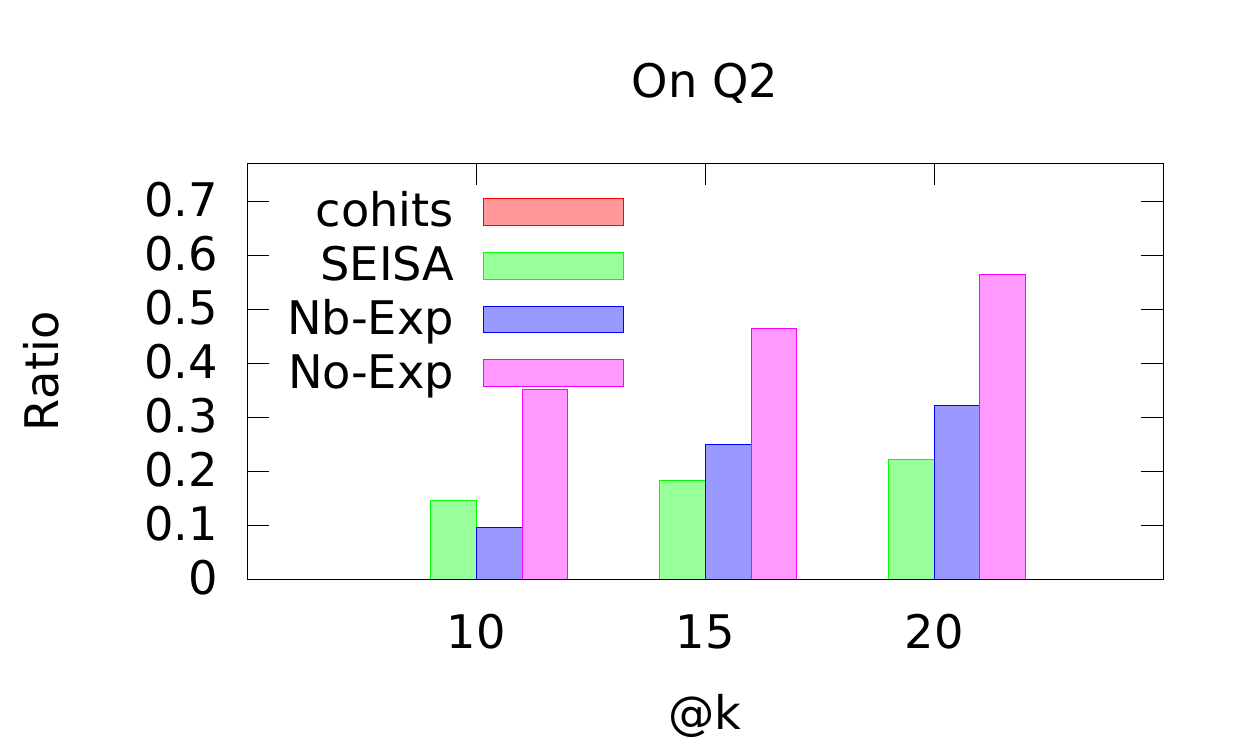}
\vspace{-7mm}
\caption{\small{Ratio $@k$}}
\vspace{-5mm}
\label{fig:ratio}
\end{figure}

\begin{figure}
\centering
\includegraphics[scale=0.325]{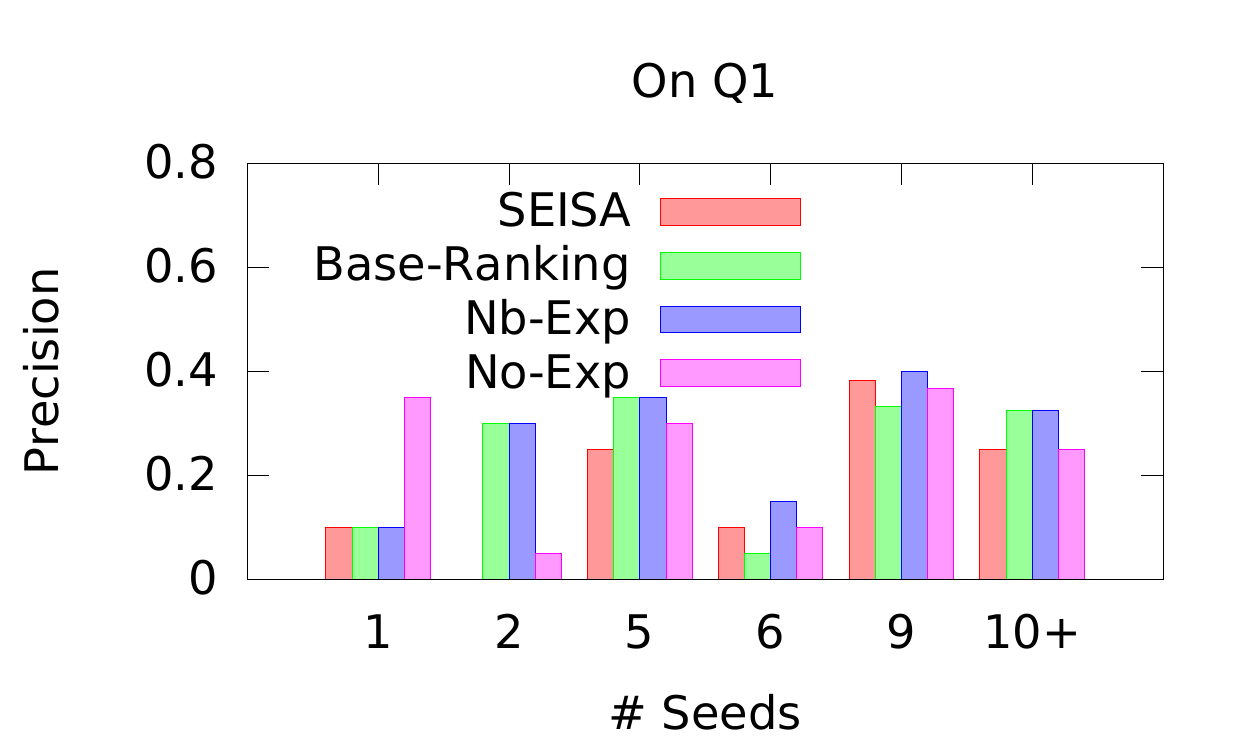}
\includegraphics[scale=0.325]{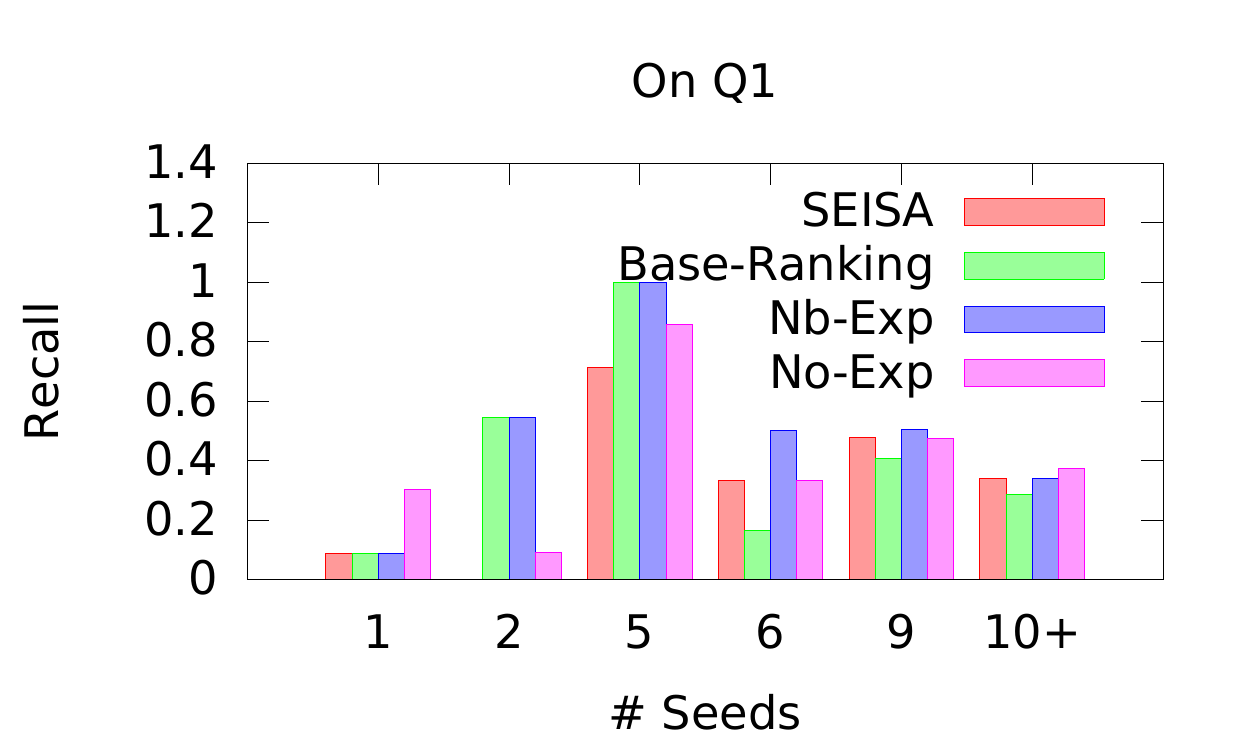}
\vspace{-7mm}
\caption{\small{Average Precision and Average Recall $@20$ on $Q_1$}}
\vspace{-5mm}
\label{fig:wiki}
\end{figure}

\subsubsection{Competitors} 
We compare with the following solutions.
\begin{itemize}
\small
\item \textbf{Decomposition-then-intersection (Int-Pro)}~ 
This is the high-precision baseline (Int-Pro for short) which we introduced in the introduction. In this baseline, we rank the results by their sum of co-occurrence frequency with the shorter concepts in $\mathbf{C}_q$.

\item \textbf{Co-Hits (PageRank)}~
Our problem can been seen as a problem of set expansion given the seed entities. Hits, PageRank and their variations have been adopted in some set expansion works such as SEAL~\cite{seal}. Since we can construct bipartite graph consisting of entities and concepts, the most appropriate one to compare is the generalized Co-Hits algorithm~\cite{cohits}.

\item \textbf{SEISA}~
SEISA~\cite{seisa} is a state-of-the-art set expansion algorithm using web lists. Here we implement its {\it Static Thresholding Algorithm} using the entity intersection of the shorter concepts of the query as the seed. Since we do not have a big web list corpus, we use Probase here instead.

\item \textbf{Base-Ranking}~
This is the baseline ranking proposed in Section 4, which directly ranks entities without additional information from the expanded concepts.

\item \textbf{Nb-Exp}~
This is our approach using Naive-Bayes model to expand concepts.

\item \textbf{No-Exp}~
This is our approach using Noisy-Or model to expand concepts.

\end{itemize}

\subsubsection{Metrics}
We have two major components: the concept expansion part and the ranking aggregation part to evaluate.
As for the concept expansion, our purpose is to evaluate how many new answer entities we can find to improve the recall. Let $n^{*}$ be the number of entities appearing in the top-$k$ answers but not in $\cap_{c\in \mathbf{C}_q}e(c)$. Then we define $ratio@k$ to quantify the ability of our solution to find new entities:
\begin{equation}
\small
ratio@k=\frac{n^{*}}{|\cap_{c\in \mathbf{C}_q}e(c)|+1}
\end{equation}   
To every query data set, we use average $ratio@k$ to evaluate the concept expansion part.

As for ranking aggregation, we report Precision$@k$ and Recall$@k$. Let the number of the entities in the ranked list be $n_r$, and the number of the entities in the ground truth be $n_g$. Precision$@k$ is the number of the correct entities that are among the top-$k$ in the ranked list divided by $n_r$. Recall$@k$ is the number of the correct entities that are among the top-$k$ in the ranked list divided by $n_g$. 

\subsection{Results}

\begin{figure}
\centering
\includegraphics[scale=0.325]{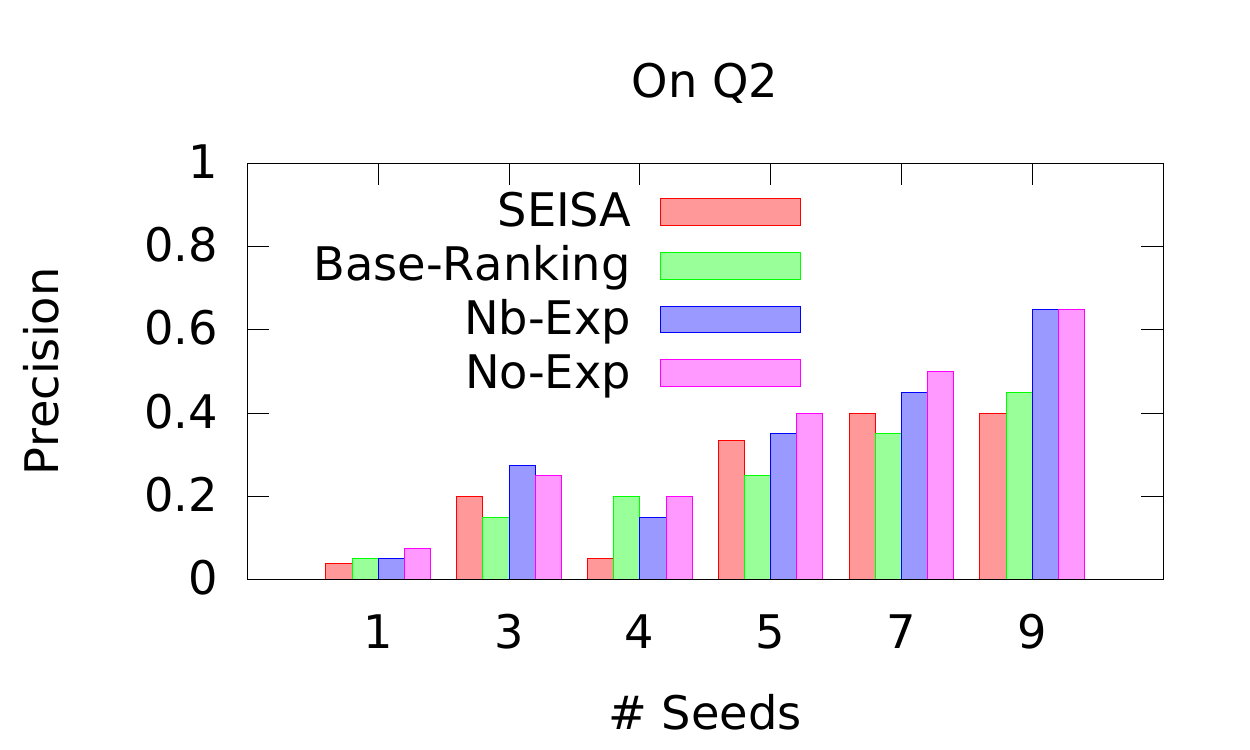}
\includegraphics[scale=0.325]{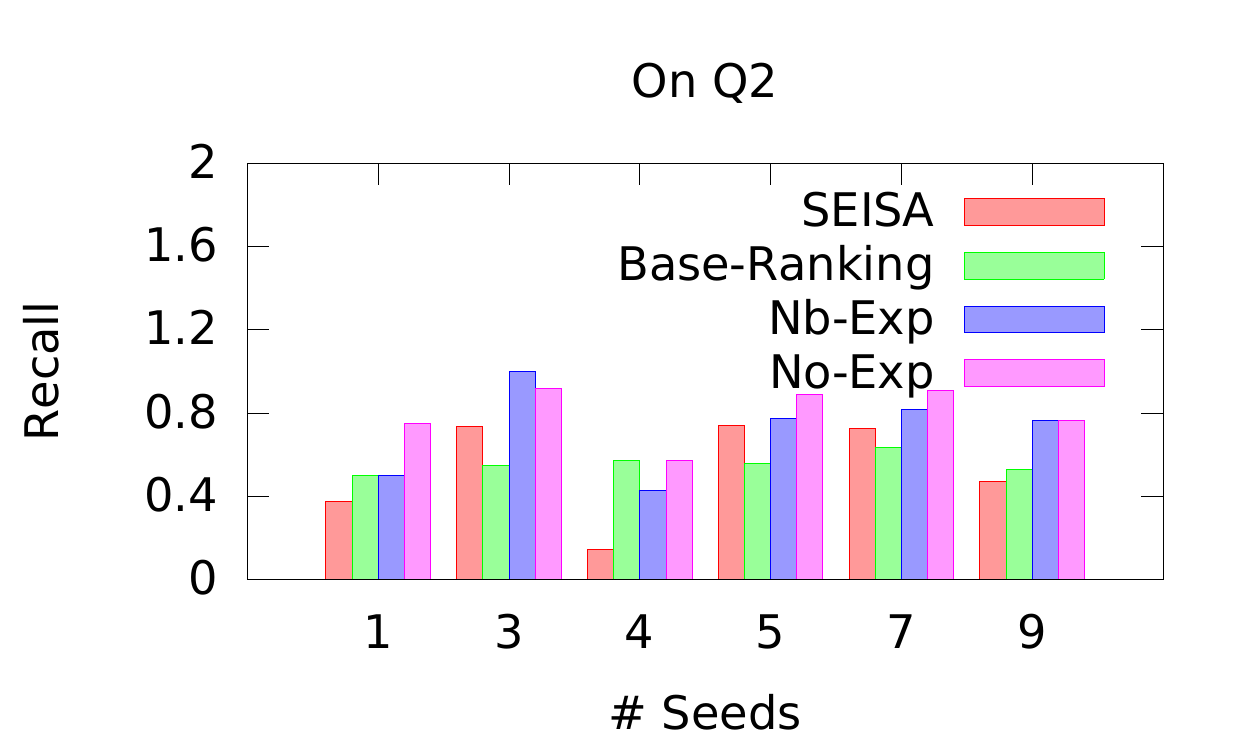}
\vspace{-7mm}
\caption{\small{Average Precision and Average Recall $@20$ on $Q_2$}}
\vspace{-5mm}
\label{fig:com}
\end{figure}

\nop{
\begin{table*}
\centering
\small{
\small{
\caption{Top Entities of Query: Largest Chinese State-owned Bank}
\begin{tabular}{|c|c|c|c|} \hline
Int-Pro & SEISA & No-Exp & GT from Wiki\\ \hline
commercial bank of china & allahabad bank & commerical bank of china & agricultural bank of china \\ \hline
industrial & andhra bank & industrial & bank of china \\ \hline
~& comerica & bank of china& bank of communications \\ \hline
~& corporation bank & bank of communications &china citic bank \\ \hline
~& santander & china construction bank & china constructio bank \\ \hline
~& commercial bank of china & agricultural bank of china & china development bank \\ \hline
~& rbs & china development bank & exim bank of china \\ \hline
~& commerzbank & \tabincell{c}{china international capital\\ corporation limited }& hua xia bank\\ \hline
~& goldman sachs group inc & china merchants bank &industrial\\ \hline
~& pnb & china minsheng banking corp &people's bank of china \\ \hline
~& pnc & export import bank of china &postal savings bank of china\\ \hline
\end{tabular}
}
}
\end{table*} 
}

\paragraph*{Precision and Recall}
In Figure~\ref{fig:wp}, we report precision and recall $@5$, $@10$, $@15$, and $@20$ (from left to right) of
our solutions with the comparison with all competitors.
On $Q_2$, we find that our No-Exp approach always performs best both on precision and recall. It also shows that in this task Noisy-Or model is better than Naive-Bayes Model. Set expansion methods do not perform as well as our approaches in this data set.
On $Q_1$, we can see that the whole performance is not as good as the performance in the result of $Q_2$. However, our Nb-Exp still performs the best on both precision and recall which reveals the effectiveness of our framework.

\nop{Evaluating the effect of different number of seeds, we can find in Figure~\ref{fig:wiki} that except the situation of only one seed existing our Nb-Exp approach performs best. When there is only one seed, No-Exp always performs best.}

\paragraph*{Influence of $\#seeds$}
The number of entities in the intersection of the shorter concepts determines the performance of set expansion algorithm.
We denote this number of $\# seeds$.
Hence, we evaluate precision and recall as the function of $\# seeds$. The results on $Q_1$ and $Q_2$ are shown in Figure~\ref{fig:wiki} and~\ref{fig:com} respectively. We can see that on $Q_2$, one of No-Exp and Nb-Exp always performs best on both precision and recall. When there is only a few seeds, set expansion algorithm is not a good choice. It is reasonable because there is not enough signal to expand concepts or entities.        

\paragraph*{Effectiveness of expansion}
Note that we propose concept expansion and seed instances generation to improve the recall. 
We find that our expansion efforts can find new entities outside of entity intersections of short concepts in Figure~\ref{fig:ratio}. The results show that our No-Exp method can always find more new entities on both of the two data sets than other methods. 
Note that Base-Ranking is only a raking solution which can not find new entities and thus is not tested here.

\subsection{Case Study}
We further give case studies to show the rationality of our models and solutions. In Table~1 and Table~\ref{tab:2}, we show the results of the query \textsl{`Top American Public University'} and \textsl{`Asian Arab Oil-exporting Country'}. 
We give results of our solution as well as the competitors including SEISA and Int-Pro.
The results show that we have a much higher recall than Int-Pro, and higher precision and recall than SEISA.
These suggest the effectiveness of our approach. 

\begin{table}
\centering
\scriptsize
\vspace{-4mm}
\caption{ \small{Top Entities of Query: Top American Public University}}
\begin{tabular}{cccc}
\toprule
Int-Pro & SEISA & No-Exp & GT from Wiki\\ 
\midrule
berkeley&berkeley&berkeley&\tabincell{c}{university of \\california berkeley}\\ \hline
 &cambridge&ucla&\tabincell{c}{university of california \\los angeles}\\  \hline
 &louisville&michigan&\tabincell{c}{university of virginia\\ charlottesville}\\ \hline
 &\tabincell{c}{harvard university}&\tabincell{c}{university of\\ washington}&\tabincell{c}{university of michigan}\\ \hline
 &\tabincell{c}{new york university}&madison&\tabincell{c}{university of north \\carolina chapel hill}\\
\bottomrule
\end{tabular}
\label{tab:1}
\end{table} 

\begin{table}
\centering
\scriptsize
\vspace{-4mm}
\caption{\small{Top Entities of Query: Asian Arab Oil-exporting Country}}
\begin{tabular}{cccc}
\toprule
Int-Pro&SEISA&No-Exp&GT from Wiki\\
\midrule
dubai&kuwait&dubai&saudi arabia \\ \hline
iran&uae&saudi arabia&iraq \\ \hline
kuwait&bahrain&kuwait&uae \\ \hline
saudi arabia&saudi arabia&iraq&kuwait \\ \hline
uae&qatar&bahrain&qatar \\ \hline
~&lebanon&oman&oman \\ \hline
~&iran&qatar&bahrain \\
\bottomrule
\end{tabular}
\label{tab:2}
\end{table}

\section{Conclusion}

This paper studies the long concept query (LCQ), that a user provides a long concept consisting of a single head with more than one modifiers and obtains in return top-$k$ entities which are the k most typical instances of the query. Such entity acquisition queries are increasingly important and common in people's daily life. However, due to the limited concept coverage of the knowledge base and the data sparsity led by the patterns building the knowledge base, it is difficult to deal with this kind of query well, specially on recall. This paper provides a query processing method based on a conceptual taxonomy which aims to improve the recall of LCQ while maintaining a high precision. Our proposed method detects some semantically related concepts based on a probabilistic framework to populate the results based on a learning-based aggregation approach considering both entity orderings and pairwise ordering constraints. Our evaluation results with real data sets show that the results of the queries processed with this new method is significantly higher than that of existing solutions.
\nop{
This paper studies the long concept query, a technique which is increasingly important and common in people's daily life. Due to the lack of the structural information of the search engine and the limited concept coverage and data sparsity of the knowledge base, this kind of queries can not be dealt with well, specially always has low recall. We leverage the new opportunities brought by many web-scale conceptual taxonomies with rich isA concept-instance relationships. Specifically, we proposed two probabilistic methods to expand concepts based on the conceptual taxonomy to populate the results. We also propose a learning-based algorithm to aggregate different information from the original concepts and also concepts expanded which aims to improve the recall of the query whiling maintaining a high precision. The effectiveness of our approaches has been validated using evaluations with real-life data.}

\newpage
\bibliographystyle{aaai}
\bibliography{reference}
\end{document}